\documentclass[aps,prl,twocolumn,groupedaddress]{revtex4}

\usepackage{graphicx}
\usepackage{latexsym}

\begin{document}


\title{Determination of the Newtonian Gravitational Constant Using Atom Interferometry}


\author{G. Lamporesi$^1$}
\author{A. Bertoldi$^1$}
\author{L. Cacciapuoti$^2$}
\author{M. Prevedelli $^3$}
\author{G.M. Tino$^1$}
\email[]{Guglielmo.Tino@fi.infn.it}
\affiliation{$^1$Dipartimento di Fisica and LENS, Universit\`a di
Firenze - INFN Sezione di Firenze, Via Sansone 1, 50019 Sesto
Fiorentino, Italy\\
$^2$ESA Research and Scientific Support Department, ESTEC,
Keplerlaan 1 - P.O. Box 299, 2200 AG Noordwijk ZH, The Netherlands\\
$^3$Dipartimento di Chimica Fisica e Inorganica, Universit\`a di
Bologna, V.le del Risorgimento 4, 40136 Bologna, Italy}


\date{\today}

\begin{abstract}
We present a new measurement of the Newtonian gravitational constant
$G$ based on cold atom interferometry. Freely falling samples of
laser-cooled rubidium atoms are used in a gravity gradiometer to
probe the field generated by nearby source masses. In addition to
its potential sensitivity, this method is intriguing as gravity is
explored by a quantum system. We report a value of
$G=6.667\cdot10^{-11}\,\textrm{m}^3\,\textrm{kg}^{-1}\,\textrm{s}^{-2}$,
estimating a statistical uncertainty of
$\pm0.011\cdot10^{-11}\,\textrm{m}^3\,\textrm{kg}^{-1}\,\textrm{s}^{-2}$
and a systematic uncertainty of
$\pm0.003\cdot10^{-11}\,\textrm{m}^3\,\textrm{kg}^{-1}\,\textrm{s}^{-2}$.
The long-term stability of the instrument and the signal-to-noise
ratio demonstrated here open interesting perspectives for pushing
the measurement accuracy below the 100 ppm level.
\end{abstract}


\maketitle

The Newtonian constant of gravity $G$ is one of the most measured
fundamental physical constants and at the same time the least
precisely known. Improving the knowledge of $G$ has not only a pure
metrological interest, but is also important for the key role that
it plays in theories of gravitation, cosmology, particle physics, in
geophysical models, and astrophysical observations. However, the
extreme weakness of the gravitational force and the impossibility of
shielding the effects of gravity make it difficult to measure $G$,
while keeping systematic effects well under control. Many of the
measurements performed to date are based on the traditional torsion
pendulum method \cite{TorsionPendulum}, direct derivation of the
historical experiment performed by Cavendish in 1798. Recently, many
groups have set up new experiments based on different concepts and
with completely different systematics: a beam-balance system
\cite{Schlamminger06}, a laser interferometry measurement of the
acceleration of a freely falling test mass \cite{Schwarz98},
experiments based on Fabry-Perot or microwave cavities \cite{Ni99,
Kleinevoss99}. However, the most precise measurements available
today still show substantial discrepancies, limiting the accuracy of
the 2006 CODATA recommended value for $G$ to 1 part in $10^4$. From
this point of view, the realization of conceptually different
experiments can help to identify still hidden systematic effects and
therefore improve the confidence in the final result.

Cold-atom interferometry has demonstrated outstanding performances
for the measurement of tiny rotations and accelerations and it is
widely used for many applications: precision measurements of gravity
\cite{Peters99}, gravity gradient \cite{McGuirk02}, and rotation of
the Earth \cite{Gustavson97, Canuel06}, but also tests of the
Einstein's weak equivalence principle \cite{Fray04}, tests of the
Newton's law at short distances \cite{Ferrari06}, and measurement of
fundamental physical constants \cite{Clade06, Mueller06}.
Applications of these techniques for fundamental physics experiments
in space are under study \cite{Tino07}.

In this paper, we present a new determination of the Newtonian
constant of gravity based on cold-atom interferometry. An atomic
gravity gradiometer is used to measure the differential acceleration
experienced by two freely falling samples of laser-cooled rubidium
atoms under the influence of nearby tungsten masses. The measurement
is repeated in two different configurations of the source masses and
modeled by a numerical simulation. From the evolution of the atomic
wavepackets and the distribution of the source masses, we evaluate
the expected differential acceleration, having G as unique free
parameter. A value for the Newton's constant of gravity is finally
extracted by comparing experimental data and numerical simulations.
Proof-of-principle experiments with similar schemes using lead
masses were already presented in \cite{Bertoldi06, Fixler07}. In the
present work, specific efforts have been devoted to the control of
systematic effects related to atomic trajectories, positioning of
source masses, and stray fields. In particular, the high density of
tungsten is crucial in our experiment to compensate for the Earth
gravity gradient. In both configurations of the source masses, atom
interferometers can therefore be operated in spatial regions where
the overall acceleration is slowly varying and the sensitivity of
the measurement to the initial position and velocity of the atoms
strongly reduced.

Our atom interferometer uses light pulses to stimulate
$^{87}\textrm{Rb}$ atoms on the two-photon Raman transition between
the hyperfine levels $F=1$ and $F=2$ of the ground state
\cite{Kasevich91b}. The light field is generated by two
counter-propagating laser beams with wave vectors $\mathbf{k}_1$ and
$\mathbf{k}_2\simeq-\mathbf{k}_1$ aligned along the vertical axis.
Atoms interact with the Raman lasers on the three-pulse sequence
$\pi/2-\pi-\pi/2$ which splits, redirects, and recombines the atomic
wavepackets. At the end of the interferometer, the probability of
detecting the atoms in the state $F=2$ is given by
$P_2=1/2\cdot(1-\cos\Phi)$, where $\Phi$ represents the phase
difference accumulated by the wavepackets along the two
interferometer arms. In the presence of a gravity field, atoms
experience a phase shift
$\Phi=(\mathbf{k}_1-\mathbf{k}_2)\cdot\mathbf{g}\,\,T^2$ depending
on the local gravitational acceleration $\mathbf{g}$
\cite{Kasevich92}. The gravity gradiometer consists of two absolute
accelerometers operated in differential mode. Two spatially
separated atomic clouds in free fall along the same vertical axis
are simultaneously interrogated by the same Raman beams to provide a
measurement of the differential acceleration induced by gravity on
the two samples.

\begin{figure}
\begin {center}
\includegraphics[width=0.4\textwidth]{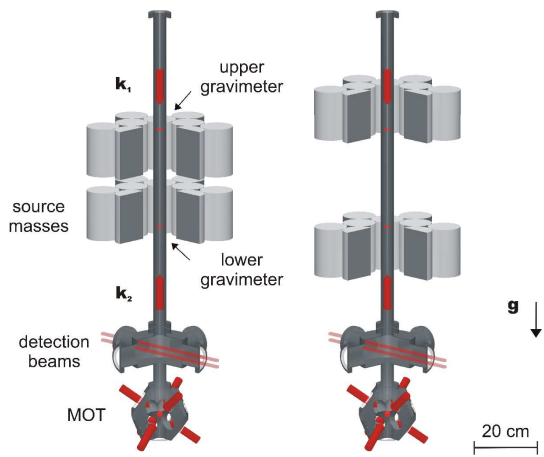}
\caption{\label{setup}Schematic of the experiment showing the
gravity gradiometer set-up with the Raman beams propagating along
the vertical direction. During the $G$ measurement, the position of
the source masses is alternated between configuration $C_1$ (left)
and $C_2$ (right).}
\end{center}
\end{figure}

Figure~\ref{setup} shows a schematic of the experiment. The gravity
gradiometer set-up and the configurations of the source masses
($C_1$ and $C_2$) used for the $G$ measurement are visible. Further
details on the atom interferometer apparatus and the source masses
assembly can be found in \cite{Bertoldi06,Lamporesi07}.
A magneto-optical trap (MOT) with beams oriented in a 1-1-1
configuration collects rubidium atoms from the background vapor of
the chamber at the bottom of the apparatus. Using the moving
molasses technique, the sample is launched vertically along the
symmetry axis of the vacuum tube and cooled down to a temperature of
$2.5\,\mu\textrm{K}$. The gravity gradient is probed by two atomic
clouds moving in free flight along the vertical axis of the
apparatus and simultaneously reaching the apogees of their ballistic
trajectories at 60 cm and 90 cm above the MOT. Such a geometry,
requiring the preparation and the launch of two samples with high
atom numbers in a time interval of about $100\;\textrm{ms}$, is
achieved by juggling the atoms loaded in the MOT: A first cloud of
$10^9$ atoms is launched to 80 cm above the MOT; during its
ballistic flight, $5\cdot10^8$ atoms are loaded and launched to 90
cm; finally, the first cloud is recaptured in the MOT and launched
again to a height of 60 cm. Shortly after this sequence, the two
atomic samples are velocity selected and prepared in the $(F=1,
m_F=0)$ level. The interferometers take place at the center of the
vertical tube shown in Fig.~\ref{setup}. In this region, surrounded
by a system of two $\mu$-metal shields (76 dB attenuation factor in
the axial direction), a uniform magnetic field of
$\sim250\,\textrm{mG}$ along the vertical direction defines the
quantization axis. The field gradient along this axis is lower than
$5\cdot10^{-5}\,\textrm{G/mm}$. The population of the ground state
is measured in the chamber placed just above the MOT by selectively
exciting the atoms on the $F=1,\,2$ hyperfine levels and detecting
the light-induced fluorescence emission. We typically detect $10^5$
atoms on each rubidium sample after the interferometer sequence.

The Raman beams are generated by two extended-cavity diode lasers
and amplified in a tapered amplifier. An optical phase-locked loop
keeps their frequency difference in resonance with the transition
between the two hyperfine levels of the $^{87}$Rb ground state
($\sim$ 6.8 GHz) and stabilizes their relative phase to about 100
mrad (rms) in a 5 Hz$-$10 MHz bandwidth \cite{Cacciapuoti05}. The
two co-propagating beams enter the vacuum system from the bottom of
the MOT cell with the same linear polarization, travel along the
axis of the vertical tube and, after crossing a quarter-wave plate,
are reflected back producing a \emph{lin}$\bot$\emph{lin}
polarization scheme. Their frequency difference is ramped down with
continuous phase during the experiment cycle to compensate for the
Doppler effect and ensure that freely falling atoms remain resonant
with the laser light. In this way, only one pair of
counterpropagating laser beams with crossed linear polarizations is
able to stimulate the atoms on the two-photon transition. The
three-pulse interferometer has a duration of $2T=320\,\textrm{ms}$.
The $\pi$ pulse lasts 48 $\mu$s and occurs $5\,\textrm{ms}$ after
the atomic clouds reach their apogees. Therefore, during the
$\pi/2-\pi-\pi/2$ pulse sequence, atomic trajectories are almost
symmetric with respect to the apogees, allowing a better control of
systematic effects induced by spurious magnetic fields.

The source masses \cite{Lamporesi07} are composed of 24 tungsten
alloy (INERMET IT180) cylinders, for a total mass of about 516 kg.
They are positioned on two titanium platforms and distributed in
hexagonal symmetry around the vertical axis of the tube (see
Fig.~\ref{setup}). Each cylinder is machined to a diameter of 100.00
mm and a height of 150.20 mm after a hot isostatic pressing
treatment applied to compress the material and reduce density
inhomogeneities. The internal structure of the material has been
analyzed using different methods: ultrasonic tests, microscope
analysis, and surface studies. Finally, a destructive test has been
performed to characterize the density distribution. We have measured
a maximum density variation of $2.6 \cdot 10^{-3}$ and an average
density inhomogeneity of $6.6 \cdot 10^{-4}$ over volume samples of
about 1/40 of the whole cylinder volume.

To evaluate the stability of the gravity gradiometer, we performed a
5-hour measurement run in which the differential phase shift was
continuously monitored. Each atom interferometer in the gravity
gradiometer measures the local acceleration with respect to the
common reference frame identified by the wavefronts of the Raman
lasers. Therefore, even if the phase noise induced by vibrations on
the retro-reflecting mirror completely washes out the atom
interference fringes, the signals simultaneously detected on the
upper and lower accelerometer remain coupled and preserve a fixed
phase relation. As a consequence, when the trace of the upper
accelerometer is plotted as a function of the lower one,
experimental points distribute along an ellipse. The differential
phase shift is then obtained from the eccentricity and the rotation
angle of the ellipse fitting the data \cite{Foster02}. We have
evaluated the Allan variance of the differential phase shift and
verified that it decays as the inverse of the integration time,
showing the typical behavior expected for white noise. The
instrument has a sensitivity of $140\,\textrm{mrad}$ at
$1\,\textrm{s}$ of integration time, corresponding to a sensitivity
to differential accelerations of $3.5\cdot10^{-8}\,g$ in
$1\,\textrm{s}$. These performances are compatible with a numerical
simulation assuming quantum-projection-noise-limited detection with
$10^5$ atoms.

\begin{figure}
\begin{center}
\includegraphics[width=0.35\textwidth]{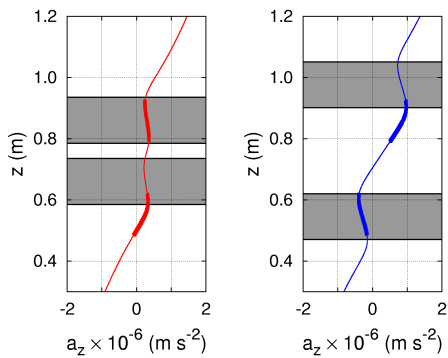}
\caption{\label{acceleration}Acceleration along the symmetry axis of
the system in configuration $C_1$ (left) and $C_2$ (right). The
simulation includes the effect of the source masses and the Earth
gravity gradient. The spatial extension of the atomic trajectories
for the upper and lower interferometers is indicated by the thick
lines. The shaded regions show the positions of the source masses.}
\end{center}
\end{figure}

\begin{figure}
\begin{center}
\includegraphics[width=0.38\textwidth]{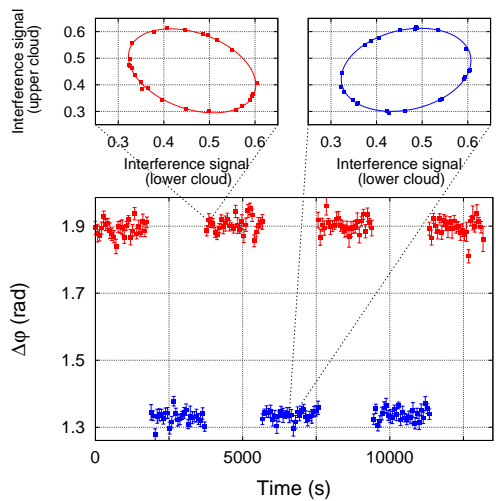}
\caption{\label{data}Typical data set showing the modulation of the
differential phase shift measured by the atomic gravity gradiometer
when the distribution of the source masses is alternated between
configuration $C_1$ (red squares) and $C_2$ (blue squares). Each
phase measurement is obtained by fitting a 24-point scan of the atom
interference fringes to an ellipse.}
\end{center}
\end{figure}

The Newtonian gravitational constant has been obtained from a series of gravity
gradient measurements performed by periodically changing the vertical position
of the source masses between configuration $C_1$ and $C_2$, with the atoms
always launched along the same trajectories. Figure~\ref{acceleration} shows
the acceleration along the symmetry axis for both configurations. Because of
the high density of tungsten, the gravitational field produced by the source
masses is able to compensate for the Earth gravity gradient. Operating the
interferometers close to these stationary points reduces the uncertainty on $G$
due to the knowledge of the atomic positions by two orders of magnitude.
A typical data sequence used for the measurement of $G$ is shown in
Fig.~\ref{data}. Each phase measurement is obtained by fitting a 24-point scan
of the atom interference fringes to an ellipse. The modulation on the
differential phase shift measured by the gravity gradiometer can be resolved
with a signal-to-noise ratio of 180 after about one hour. Experimental data
have been collected during two separate measurement runs performed in February
and May 2007.

The differential nature of the measurement ensures an efficient
rejection of many systematic errors: Firstly, each gravity gradient
measurement is highly insensitive to phase shifts appearing in
common mode on the two interferometers \cite{McGuirk02}; in
addition, the differential measurement performed by alternating the
source masses between configuration $C_1$ and $C_2$ removes effects
not depending on the masses position. A budget of the systematic
errors affecting the value of $G$ is presented in
Table~\ref{systematics}.
\begin{table}
\begin{center}
\begin{tabular}{lc}
\hline\hline
Systematic effect & $\Delta G/G$ ($\times 10^{-4}$) \\
\hline
Radial position & 1.2\\
Vertical position in $C_1$ & 2.7\\
Vertical position in $C_2$ & 2.1\\
Cylinders mass & 0.9 \\
Cylinders density inhomogeneity & 0.21\\
Support platforms mass & 0.8\\
Initial position of the atomic clouds & 0.18\\
Initial velocity of the atomic clouds & 2.3\\
Gravity gradient & 1\\
Stability of the on-axis B-field & 0.3\\
Stability of the launch direction & 0.6\\
\hline
Total & 4.6\\
\hline\hline
\end{tabular}
\caption{\label{systematics}Error budget of the systematic shifts
affecting the $G$ measurement.}
\end{center}
\end{table}
Each effect is taken into account in our numerical simulation and
its contribution evaluated on the basis of the measurements
performed on the relevant parameters. Positioning errors account for
uncertainties in the position of the 24 tungsten cylinders along the
radial and vertical direction, both in configuration $C_1$ and
$C_2$. Density inhomogeneities of the source masses have been
modeled. An upper bound of the systematic uncertainty on $G$ has
been evaluated on the basis of the maximum density variation
measured during the characterization of the tungsten cylinders. The
contribution of the platforms holding the source masses has been
included as well. The velocity of the atomic clouds and their
position at the time of the first interferometer pulse have been
calibrated performing time-of-flight measurements and detecting the
atoms when crossing a horizontal light sheet twice, once on the way
up and once on the way down. The knowledge of the gravity gradient
is also important to identify the best positions for the apogees of
the two atomic clouds with respect to the source masses (see
Fig.~\ref{acceleration}). Finally, special care has to be taken of
all the systematic errors that could depend on the source masses
distribution. For instance, the presence of an inhomogeneous
magnetic field along the vertical axis depending on the position of
the source masses can introduce a systematic phase shift affecting
the $G$ measurement. However, the low magnetic susceptibility of the
tungsten alloy and the two $\mu$-metal shields surrounding the
interferometer region ensure an excellent control of stray fields
and reduce possible systematic errors below the measurement
sensitivity of our apparatus. Using a tiltmeter, we have verified
that the direction of the Raman lasers, which defines the sensitive
axis of our interferometer, is stable within $1\,\mu\textrm{rad}$
and does not depend on the position of the source masses. At the
same time, the stability of the launch direction has been evaluated
and an upper bound on possible phase shifts induced by the Coriolis
acceleration on the differential measurement estimated.

\begin{figure}
\begin{center}
\includegraphics[width=0.4\textwidth]{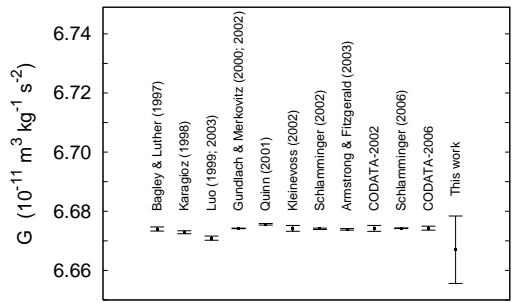}
\caption{\label{result}Our result of the Newtonian gravitational
constant compared to the most precise G measurements recently
obtained and to CODATA recommended values.}
\end{center}
\end{figure}

After an analysis of the error sources affecting our measurement, we
obtain a value of
$G=6.667\cdot10^{-11}\,\textrm{m}^3\,\textrm{kg}^{-1}\,\textrm{s}^{-2}$,
with a statistical uncertainty of
$\pm0.011\cdot10^{-11}\,\textrm{m}^3\,\textrm{kg}^{-1}\,\textrm{s}^{-2}$
and a systematic uncertainty of
$\pm0.003\cdot10^{-11}\,\textrm{m}^3\,\textrm{kg}^{-1}\,\textrm{s}^{-2}$.
Figure~\ref{result} shows this result and compares it to the most
precise G values recently obtained in different experiments. Our
measurement is consistent with the 2006 CODATA value within one
standard deviation.

We have reported a new measurement of the Newtonian constant of
gravity based on atom interferometry. This technique, completely
different from the methods used to date, explores new systematics,
globally improving the confidence in the knowledge of $G$.
Presently, our statistical error averages down reaching a value
comparable to the systematic uncertainty in about one day of
integration time. As expected, the main contribution to the
systematic error on the $G$ measurement derives from the positioning
accuracy of the source masses. This error will be reduced by about
one order of magnitude once the position of the tungsten cylinders
will be measured by a laser tracker. From the analysis of systematic
effects and the characterization of the stability of our apparatus,
we foresee interesting perspectives for pushing the measurement
accuracy below the 100 ppm level.

\begin{acknowledgments}
This work was supported by INFN (MAGIA experiment), MIUR, ESA, and
EU (under contract RII3-CT-2003-506350). G.M.T. acknowledges seminal
discussions  with M.A. Kasevich and J. Faller and useful suggestions
by A. Peters. M. Fattori, T. Petelski, and J. Stuhler contributed to
the setting up of the apparatus. We are grateful to A. Cecchetti and
B. Dulach of INFN-LNF for the design of the source masses support
and to A. Peuto, A. Malengo, and S. Pettorruso of INRIM for density
tests on W masses.
\end{acknowledgments}


\end{document}